\pdfoutput=1
\documentclass[conference]{IEEEtran}
\IEEEoverridecommandlockouts
\usepackage{cite}
\usepackage{amsmath,amssymb,amsfonts}
\usepackage{algorithmic}
\usepackage{graphicx}
\usepackage{textcomp}
\usepackage{xcolor}
\def\BibTeX{{\rm B\kern-.05em{\sc i\kern-.025em b}\kern-.08em
    T\kern-.1667em\lower.7ex\hbox{E}\kern-.125emX}}
\begin{document}

\title{Multi-level Explanation of Deep Reinforcement Learning-based Scheduling}


\author{\IEEEauthorblockN{Shaojun Zhang}
\IEEEauthorblockA{\textit{School of Computer Science} \\
\textit{The University of Sydney}\\
szha6955@uni.sydney.edu.au}
\and
\IEEEauthorblockN{Chen Wang}
\IEEEauthorblockA{\textit{CSIRO Data61} \\
Sydney, Australia \\
chen.wang@data61.csiro.au}
\and
\IEEEauthorblockN{Albert Zomaya}
\IEEEauthorblockA{\textit{School of Computer Science} \\
\textit{The University of Sydney}\\
albert.zomaya@sydney.edu.au}
}

\maketitle

\begin{abstract}
Dependency-aware job scheduling in the cluster is NP-hard. Recent work shows that Deep Reinforcement Learning (DRL) is capable of solving it. It is difficult for the administrator to understand the DRL-based policy even though it achieves remarkable performance gain. Therefore the complex model-based scheduler is not easy to gain trust in the system where simplicity is favored. In this paper, we give the multi-level explanation framework to interpret the policy of DRL-based scheduling. We dissect its decision-making process to job level and task level and approximate each level with interpretable models and rules, which align with operational practices. We show that the framework gives the system administrator insights into the state-of-the-art scheduler and reveals the robustness issue in regards to its behavior pattern.

\end{abstract}


\section{Introduction}
Job scheduling plays an important role in resource provisioning, energy-saving and cost reduction for the Cloud. However, due to increasing job complexities, e.g. the possibility of parallel execution and a large number of mutual dependencies, finding the optimal scheduler is inevitably hard. So, apart from enhancing existing heuristic algorithms, researchers turn to Deep Reinforcement Learning (DRL). Recently, various work has shown remarkable success in DRL-based scheduling. However, there is still an obstacle preventing the advantage from being turned into reality. 

Generally, the practical system welcomes simpler algorithms more often than complex ones, due to the former generates easy-to-understand behaviors. When problems occur, the system administrator can easily diagnose the causes and adjust the system accordingly. Although DRL is powerful, the administrator is cautious to hand over the system to its unpredictable black box. 

Understanding the model behaviors helps to release the issue. Most methods that seek algorithmic explanations for complex deep neural networks ~\cite{9007737,che2016interpretable,ribeiro16model,Simonyan2014DeepIC} can hardly draw the explanation from the domain-specific perspective or practical experiences. In the case of job scheduling, the operational-level questions, like how a task is selected by the scheduler for execution, why a specific job is scheduled rather than the other, and
will the scheduling policy of the DRL model align with simpler scheduling policies, are not well answered. 

In this paper, we propose a multiple-level explanation framework to address the problem. Our framework guides the system administrator to gain insights into the decision-making process of the DRL-based scheduler. First, interpretable features are extracted from the common concepts of job scheduling. They are used to train simpler models for mimicking the DRL model's decision. To balance the fidelity and interpretability, the model is dissected into multiple levels. The job level gives reasons why one job is given higher priority than the other. The task-level provides insights into how the policy is aligned with well-practiced rules. Given the understanding of DRL-based policy, we figure out the potential robustness issue within its behavior pattern. It can help to strengthen the system. In summary, the explanation framework establishes a pathway for the complex scheduler to gain the trust of administrators. Specifically, our contributions are as follows: 
\begin{enumerate}
    \item We dissect the DRL-based scheduling into job level and task level, and approximate decisions at each level with interpretable models, which together with interpretable features give explanations on how jobs are scheduled; 
    \item Based on the insight obtained from the explanation framework, we propose the \emph{node-split} method. It explores the scheduler's robustness issue by crafting perturbation on the job and helps the administrator to manage potential risks for the DRL-based policy; 
    \item We conduct extensive experiments on the state-of-the-art DRL-based scheduler and demonstrate the high fidelity in approximation and operational understanding, which is hardly achieved by algorithmic explanations. 
\end{enumerate}


\section{Background}
\subsection{DRL-based Scheduling} 
Clusters often host long-running data analytic jobs. These jobs encapsulate a large number of parallel tasks and complex inter-task dependencies. Each of them demands efficient execution and fair resource allocation. Prevalent rules or heuristic-based schedulers \cite{ferguson12jockey,ghodsi11dominant,grandl14multi,grandl2016graphene,Isard09quincy} often require a lot of human endeavor for performance tuning. As a result, the Deep Reinforcement Learning (DRL)-based scheduler becomes an alternative \cite{chen19learning,mao16resource,mao19learning,zhang20learning,zheng20machine}. Its deep neural network-based policy improves itself automatically through interaction with the system. During training, a variety of metrics can be rendered in the reward function to meet different system requirements. For example, DeepRM \cite{mao16resource} optimizes towards the best-packed resource demands while minimizing the job's slowdown. The priority dispatching rule in \cite{zhang20learning} learns to minimize the average makespan in the job-shop scenario. These applications illustrate some common patterns. First, the input and output space of the model can be extremely large: jobs are summarized to latent representations via an extra neural network, such as Graph Neural Networks (GNNs) in \cite{mao19learning,zhang20learning}, while the action is rendered in a  per-task manner \cite{mao16resource,mao19learning}. Second, the model equips with numerous non-linear layers and its behaviors are hard to explain. While achieving incredible performance improvement, the DRL-based scheduler faces a trust problem in its target area.

\subsection{Explanation of Deep Learning Models} There are three kinds of approaches to explaining the neural model. The first kind uses interpretable proxy models to mimic the behavior of the original model~\cite{che2016interpretable,Ribeiro16why}. The second kind uses a salience map or heatmap to show the sensitivity of the model output to the input features \cite{Simonyan2014DeepIC,selvaraju2017grad}. And the third applies counterfactual analysis by hypothesizing examples to obtain different predictions and understand the original prediction \cite{mothilal2020explaining}. However, purely algorithmic explanations can merely provide system administrators with operational-level explanations as rules and heuristics do. 

\section{Method}




\begin{figure}[t]
    \centering
    \includegraphics[width=0.95\linewidth]{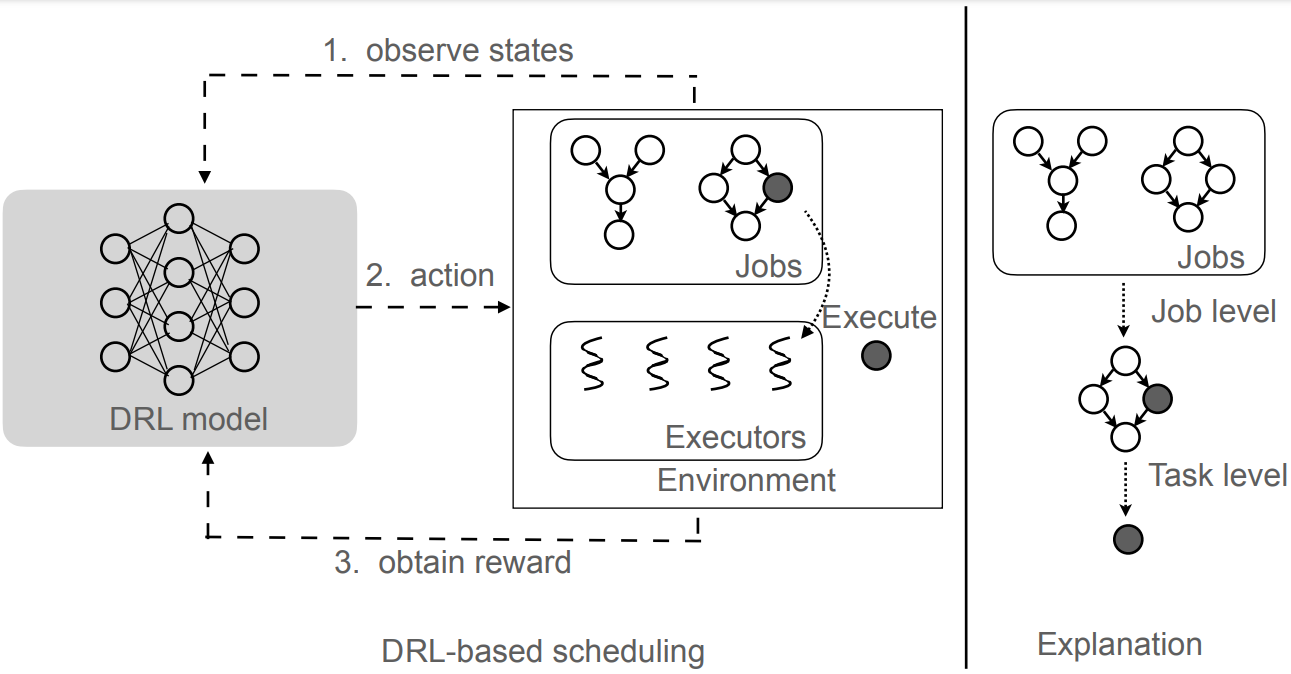}
    \caption{The DRL-based job scheduler and the multi-level explanation framework.}
    \label{fig:explanation_framework}
\end{figure}

\subsection{Problem Definition}
\paragraph{Job scheduling} Assuming the cluster hosts multiple jobs $\mathcal{J}$ and manages a number of executors $\mathcal{E}$. According to the computation pattern of data analytic frameworks such as Apache Spark or Storm, each job $J \in \mathcal{J}$ is represented as a DAG and contains a set of task nodes $\mathcal{N}_J$. Each task is replicated multiple times in a node, running on different data in parallel. The task is completed only if all replicates are executed. The scheduler assigns a task to an executor at a time, sleeps if no executor is available, and resumes itself once an executor is freed, as shown by Figure~\ref{fig:explanation_framework}. This process iterates until all jobs are completed, forming a trajectory containing thousands of time stages. At a given time stage $t$, the observed \emph{state} $S_t$ is composed of job features, e.g. the degree of parallelism of each task node and estimated per-task run time, and system status, e.g. the number of free executors currently available. The \emph{policy} $f$ encoded by DNNs maps the state to an \emph{action} $A_t$ as $f: S_t \rightarrow A_t$. Generally, each task node is given a score by $f$, and the task replicate from the node with the highest score is executed. The size of the action space is $\sum_J{|\mathcal{N}_J|}$.     

\paragraph{Multi-level explanation framework} We devise the proxy-based explanation framework to understand the scheduling policy $f$. However, the direct mimicking of its behavior with proxy models, such as sparse linear models, rules, or decision trees, has some restrictions. First, the raw job features can only provide limited operational insights over the behavior even with highly-faithful approximation. Second, the high interpretability of the proxy model and fidelity of mimic can hardly be reached at the same time for DRL.  
Hence, we formulate operational features out of concept of scheduling, and decompose the decisions of $f$ into job level $g_j$ and task level $g_n$ to align with system administrator's knowledge respectively. As shown by dash arrows in Figure ~\ref{fig:explanation_framework}, $g_j$ mimics the selection of job $J^*$ out of $\mathcal{J}$ as $g_j: S_t \rightarrow J^*$, and $g_n$ mimics the selection of task node $N*$ from $\mathcal{N}_{J^*}$ as $g_n: J^* \rightarrow N^*$. The output space of $g_j$ is $|\mathcal{J}|$, and of $g_n$ is $|\mathcal{N}_{J^*}|$. Since $|\mathcal{J}| + |\mathcal{N}_{J^*}| \ll \sum_J |\mathcal{N}_J|$, the multi-level approach can provide higher approximation with less effort.

\subsection{Interpretable Features}\label{sec:features}
Raw job and system features, such as parallelism configuration of the task and the number of executors, do not support the high-level operational-level explanation. So we derive the interpretable features $\mathcal{X}$  from common scheduling properties as follow:   

\begin{itemize}
    \item (1) The \emph{run time} and (2) \emph{task replicates}  along each job's critical path, which is composed of task nodes incurring the longest run time.
    \item (3) The \emph{run time} and (4) \emph{task replicates} for each job. 
    \item (5) The \emph{run time} and (6) \emph{task replicates} in the smallest \emph{ready} task node of each job. A ready node means all its dependencies have been resolved.
    \item (7) \emph{Locality}. Locality indicates whether the job is local to the available executor. The overhead of cold start can be costly when a job is dispatched to an executor for the first time. If locality stands, the overhead is avoided. 
\end{itemize}
Notice, that the run time is estimated by assuming sequential execution of the parallel task replicates. As the system proceeds timely, the job structure, the critical path, and the smallest ready node may be changed accordingly.

\subsection{Job Level Schedule Explanation}
Recall, at each time stage, $f$ selects a job and schedules one of its tasks for execution. To differentiate, the selected job is marked positive while the other jobs are marked negative. The positive-to-negative ratio is $1:|\mathcal{J}|-1$, contributing to a high imbalance for the dataset. $g_j$ is responsible for distinguishing the positive job from all negative jobs at all times. However, the data imbalance is hard to be handled by any interpretable model. So we devise two sub-layers for the high-faith explanation.

\paragraph{Initial approximation} The initial approximation aims to classify each job as either positive or negative based on the features outlined in Section \ref{sec:features}. A rough explanation like 'the job is selected due to some \textit{prominent features}' is conveyed. In practice, we devise the decision tree with the \emph{C4.5} algorithm \cite{quinlan93decisiontree} due to its clear interpretability.
If a tree node contains positive instances mostly and ensures small entropy, it stops further split. The prominent features for those positive instances in it are obtained by tree traversal, and jobs are selected due to them. For the nodes containing mixed positive and negative instances, however, the explanation may not hold. We provide further tuning to settle the problem.

\paragraph{Further tuning} In the tree nodes with mixed positive and negative instances, the imbalance still exists. We propose the \emph{pairing-and-comparing} method to reformulate the problem. Given a time stage $t$, there is one positive job instance $J_{p,t}$ and a set of negative instances $\{J_{n,t}\}$, where $J_{p,t}, J_{n,t} \in \mathbb{R}^{|\mathcal{X}|}$. Pairing is realized by concatenation of the positive instance with negative instance as follow: $[J_{p,t}:J_{n,t}]$ and $[J_{n,t}:J_{p,t}]$, where the symbol $:$ means concatenation of the features. To differentiate, the former is called positive-negative pair and the latter is called negative-positive pair. The aim of classification is changed from recognizing original instances to recognizing the type of pairs. Meanwhile, the explanation is changed from 'which job from $\mathcal{J}$ is selected' to 'why one job instead of the other is more likely to be scheduled'. The training model on paired instances has the following advantages. First, it focuses on jobs at the same time stage, avoiding meaningless comparing between jobs from different time stages. Then, it can solve the imbalance problem by sampling the same amount of two kinds of pairs. In practice, it can also keep data efficient by sampling a small number of (often less than 10) negative ones for concatenation. To compare, the space for original job instances is rough $|\mathcal{J}|$.

We devise the random forest (RF) model \cite{breiman01randomforest} to balance the fidelity of approximation and interpretability. The number of trees in the model is restricted to less than 20. The RF model learns to differentiate between two kinds of pairs. After classification, the jobs in each pair are decomposed and labeled either positive or negative. Then, the schedule can be explained by inter-job comparison through the decision path of the pair in the forest.  


\subsection{Task Level Schedule Explanation}
Given the selected job $J^*$, task-level approximation $g_n$ selects the task node $N_{J^*}$, whose replicate is assigned for execution. Basically, the input feature space equals the number of task nodes $|J^*|$. It is a small-scale problem that rule-based schedulers can provide not only an accurate approximation but clear interpretability. In specific, an explanation of 'how the DRL-based policy aligns to existing rules' is conveyed. There are numerous choices of rules from the community of schedule. We utilize some of the most prevalent yet simple ones as follows:
(1) shortest task node first (SNF): the task node with the least run time is scheduled;
(2) critical path-based scheduling (CPS): the task node on the critical path is scheduled sequentially, and (3) first come first serve (FCFS): the task node is selected regarding the dependencies. These rules are evaluated by the accuracy of the approximation. The one with the highest accuracy aligns with DRL-based schedule decisions the most and provides the best interpretability.

\section{Experiment}
\subsection{Experimental Settings}
\paragraph{The DRL-based scheduler} We evaluate the multi-level explanation framework on the state-of-the-art DRL-based scheduler, Decima \cite{mao19learning}. Decima adopts an end-to-end model to perceive the states and generate scheduling actions. It integrates several neural networks. Firstly, two GCNs are used to synthesize the task features to the job-level information and then to global-level information respectively. The GCNs propagate through the dependency within each job and the aggregation of information is realized with 3 fully-connected layers, whose size is 16, 8, and 1. Then a policy network is used to generate the score for each task. It is implemented with 4 fully-connected layers, whose size is 32, 16, 8, and 1. Every time there is a free executor, the model is invoked and its output score is used to select the task for execution. The task will occupy the executor’s monopoly during execution. The model is trained with the REINFORCE algorithm \cite{williams92Simple} on the TPC-H dataset for more than 3000 episodes. In each episode, 100 jobs with more than thousands of tasks are initiated to run on 15 executors. The model aims to minimize the average Job Completion Time (JCT) for them.   

\paragraph{Schedule traces for explanation} The explanation framework is trained on Decima's schedule traces. We obtain 5 traces by using Decima to manage different job sets on 5 executors. Each job set contains 30 TPC-H jobs. The statistics of each trace are listed in Table ~\ref{tab_traces}. The interpretable features in Section \ref{sec:features} are extracted for job instances at every time stage. Since there is one positive job instance at each time, the dataset is highly imbalanced. Totally, four traces are used for training the framework, while the last one is used for testing.

\paragraph{Baseline} The salience map-based explanation method shows the impacts of each input feature on the given output. We utilize the Jacobian-based Salience Map Approach (JSMA) \cite{papernot16} as the baseline. Basically, for the classification problem, a feature is marked impactful if modifications on it not only strengthen the model's decision but also decrease the weights of other candidate choices. So, the salience map is calculated by the gradients of the dominant output component w.r.t. the input space minus the summation gradients of all other components w.r.t. the input space. Notice, that this feature space is formed by the original task features: (1) the number of executors, (2) the number of source executors, (3) whether this free executor is released by this job, (4) the task replicates in the node, and (5) overall remaining task run time.    

\begin{table}[t]
\centering
\small
\caption{The statistics of Decima's schedule traces.}
\begin{tabular}{cccccc}
\hline
\multicolumn{1}{c}{Trace ID} & 1 & 2 & 3 & 4 & 5 \\ \hline
Pos. instances  & 390 & 381 & 410 & 396 & 395 \\
Neg. instances & 6298 & 6483 & 6802 & 6299 & 6913 \\ \hline
\end{tabular}
\label{tab_traces}
\end{table}


\subsection{Explanations}
\begin{figure}
    \centering
    \includegraphics[width=0.78\linewidth]{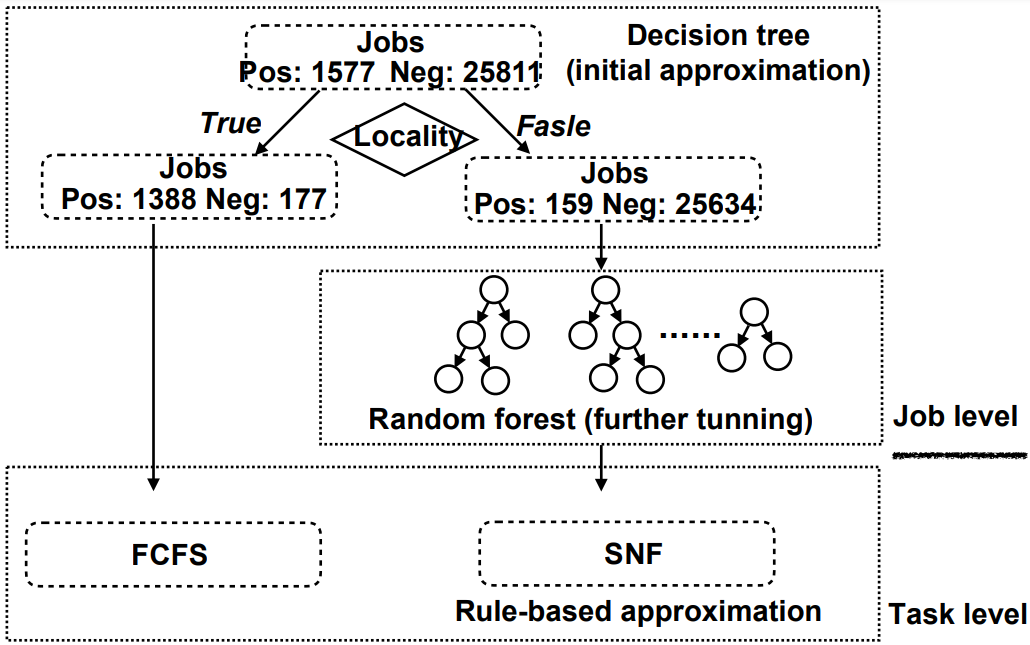}
    \caption{The multi-level explanation for DRL-based scheduler Decima. The classification of training data is included.}
    \label{fig:explanation-decima}
\end{figure}

The multi-level explanation framework after training is shown in Figure ~\ref{fig:explanation-decima}. At the job level, $g_j$ utilizes the decision tree for initial approximation and the random forest for further tuning. Then, at the task level, $g_n$ utilizes different rules for approximation under different branches from the upper level. 

\paragraph{Initial approximation} The initial approximation provides an understanding of why a job is chosen in terms of its prominent features. Jobs at this level are classified as either positive or negative with their own interpretable features. The branches that guarantee an entropy gain ratio higher than bar $\lambda$ ($>$ 0.5) are kept while others are pruned. In the tree shown by Figure~\ref{fig:explanation-decima}, the prominent feature \emph{locality} introduces two branches. During training, it achieves an entropy gain ratio of 0.73 on the dataset. In total, 1388 positive and 177 negative instances whose \emph{locality} is \emph{True} are collected by the left-side branch, while other 159 positive and 25634 negative instances whose \emph{locality} equals to \emph{False} are collected by the right-side branch. 
During testing, 347 positive and 43 negative instances are gathered in the left-side branch, while 48 positive and 6870 negative instances are in the right-side branch. The scheduler's behavior of selecting jobs in the left-hand branch can be explained by the prominent features. In all, 88.6\% jobs during training and 88.9\% jobs during testing which are local to the free executor will have their tasks scheduled. However, the right-side branch can not be purely explained with locality.  


\paragraph{Further tuning} 
Further tuning provides explanations on why one job is selected rather than another by comparison between their features when both jobs are not local to the executor. 
The jobs in the right-side tree branch after the initial approximation are approximated with the small-scale random forest. In the experiment, the number of trees is set to 15 and the maximum tree levels are set to 9. Notice, that while more trees and higher levels contribute to higher fidelity for approximation, they often increase the difficulty of explanation. To keep data efficiency, each positive instance is concatenated to 5 different negative instances to the maximum at the same time stage. In all, the model is trained on more than 700 pairs from different time stages, as shown by Figure~\ref{fig:explanation-decima}. 

\begin{figure}[t]
    \centering
    \begin{minipage}[t]{0.5\linewidth}
    \centering
    \includegraphics[width=0.85\linewidth]{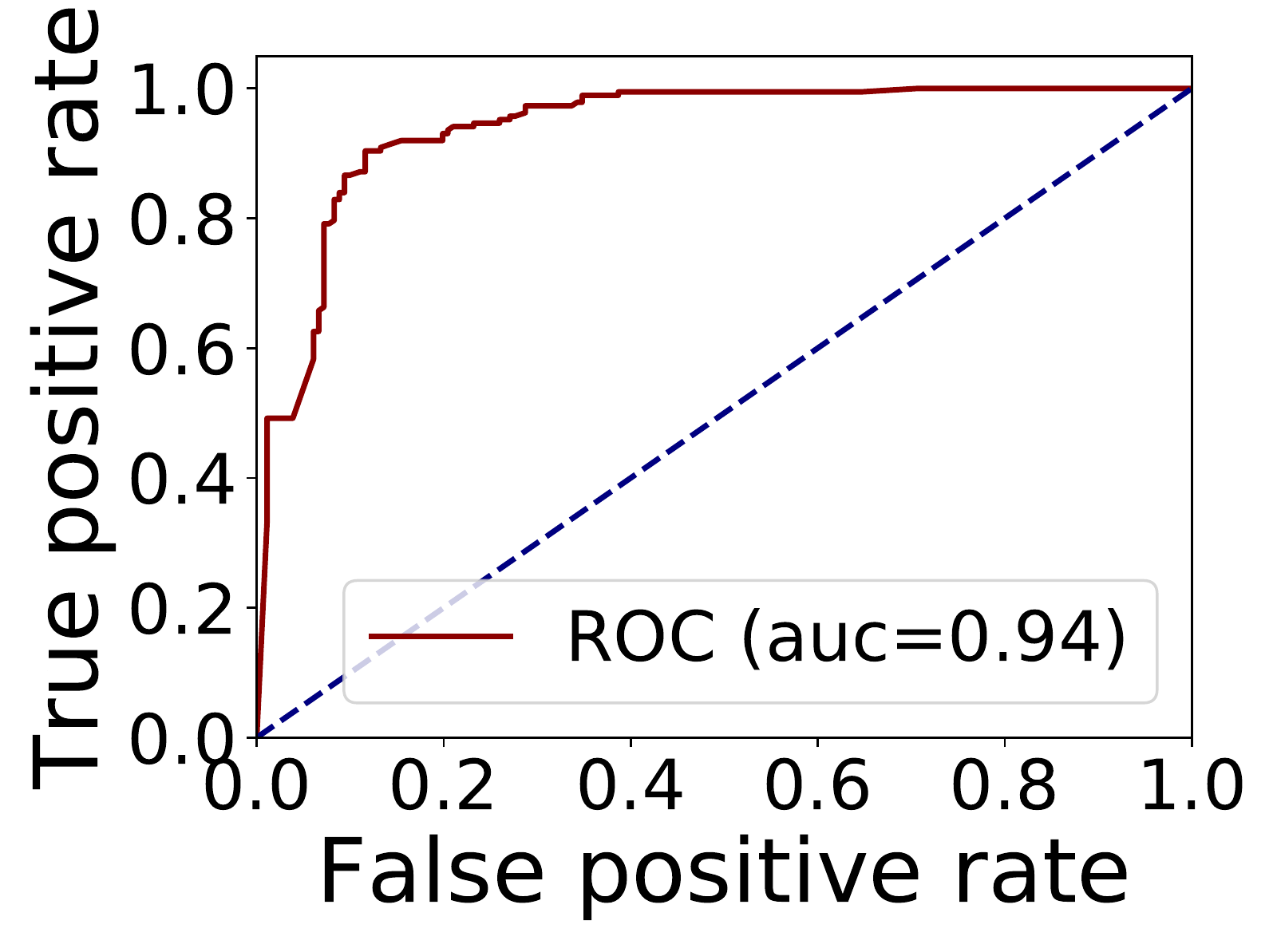}\\
    \centering{\small{\text{(a)}}}
    \end{minipage}%
    \hspace{1.2mm}
    \begin{minipage}[t]{0.46\linewidth}
    \centering
    \includegraphics[width=0.88\linewidth]{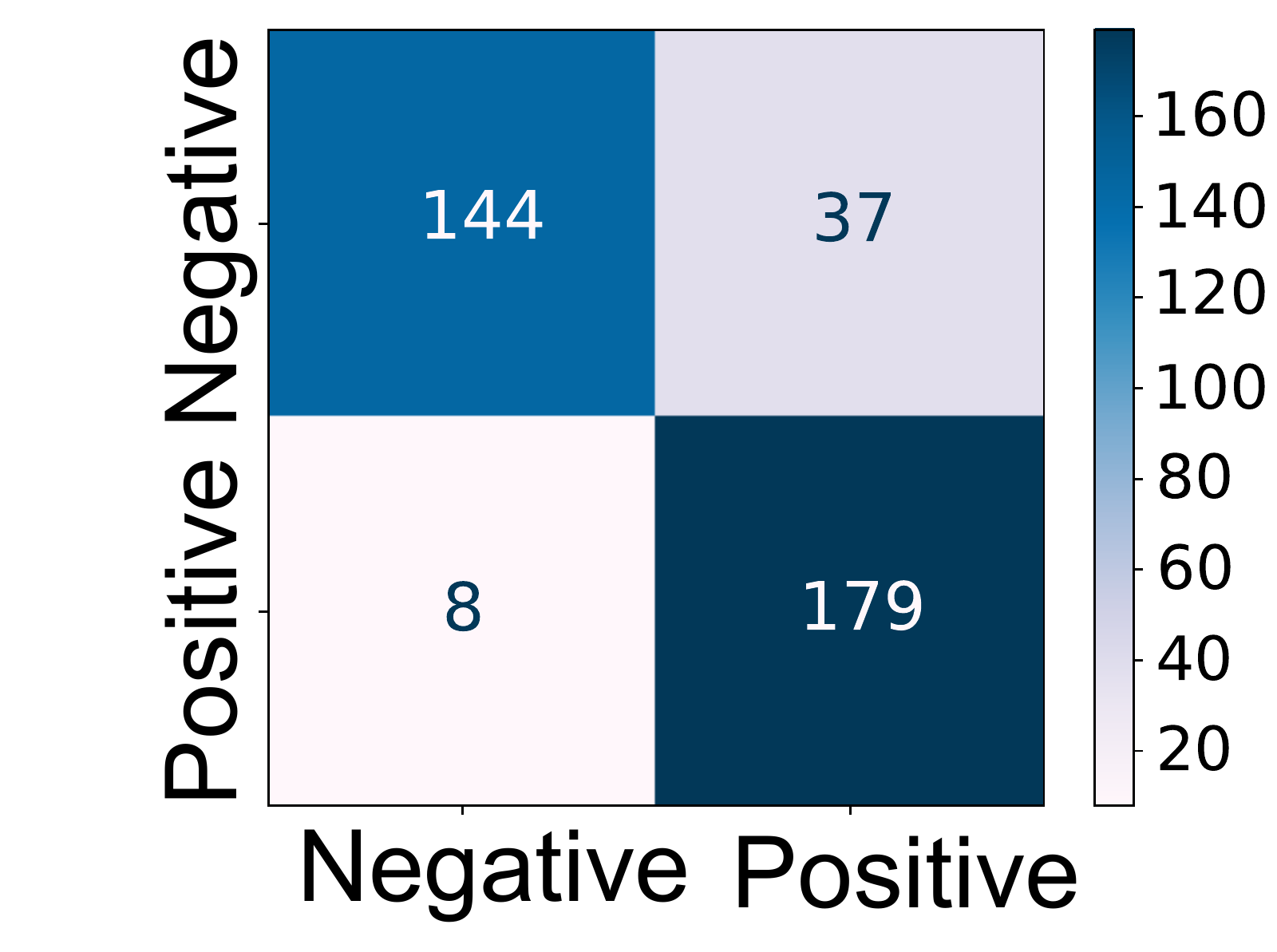}\\
    \small{\text{(b)}}
    \end{minipage}%
    \caption{The (a) ROC curve and (b) confusion matrix of positive-negative and negative-positive pairs. The positive-negative pair is noted as \textit{positive} class in the figure.}
    \label{fig_confusion_matrix}
\end{figure}

The trained RF model is testified by whether it successfully distinguishes between the two types of pairs from the test dataset. The dataset contains 187 positive-negative and 181 negative-positive pairs sampled from trace 5 in Table \ref{tab_traces} which is not seen during training. Overall, the accuracy of the classification is 0.88. The ROC curve of the binary-class classification is shown in Figure~\ref{fig_confusion_matrix}(a). The positive-negative pair is marked as the \emph{positive} class. The area under the curve (auc) accounts for 0.96 and guarantees high fidelity. The confusion matrix in Figure~\ref{fig_confusion_matrix}(b) also shows two types of pairs distinguished with high confidence.

\begin{figure}[t]
    \centering
    \includegraphics[width=0.85\linewidth]{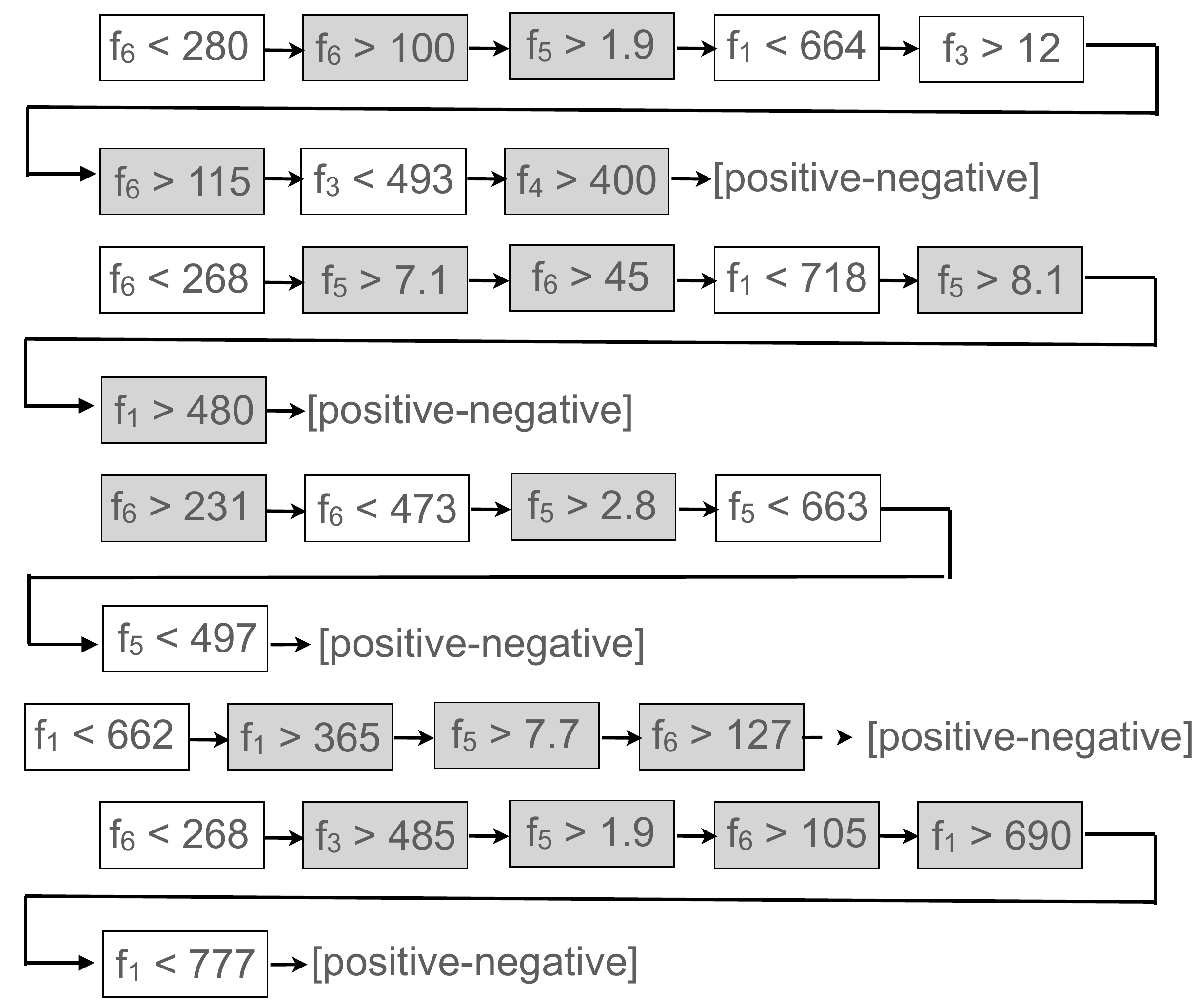}
    \caption{5 decision paths from the random forest during classifying the concatenations of jobs listed in Table \ref{tab:example_job_pair}.  The tree node includes the feature indexed in Section \ref{sec:features} and the threshold to split. The feature of the positive job is white, while of the negative is gray.}
    \label{fig:rf_explanation_examples}
\end{figure}

Decima's decisions are explained by tree traversal in the random forest. The decision path of classifying a given pair is composed of features that differentiate the concatenated jobs. We give examples of how the RF model's decision paths that differ from each other demonstrate unified logic in behaviors. As shown in Figure~\ref{fig:rf_explanation_examples}, five different decision paths are extracted during classifying the paired jobs, whose interpretable features are shown in Table \ref{tab:example_job_pair}. The tree node is split in regard to interpretable features indexed in Section \ref{sec:features}. The threshold for split is indicated in each node. To differentiate, features of the positive job are shaded white while those of the negative job are shaded gray in Figure~\ref{fig:rf_explanation_examples}. The output of the path points to the class of pair, either positive-negative or negative-positive. From the paths, we have some direct observations. First, only a small group of features are essential to discriminate the type of pairs. In other words, the DRL-based scheduler makes decisions by considering these features more often. Second, the same feature of the paired jobs is prone to be included in the same path. For example, the task replicates of the smallest ready task node ($f_6$) of both jobs exist in the 1st, 2nd, 3rd, and 5th tree path in Figure~\ref{fig:rf_explanation_examples}. The value of such a feature provides insights into why one job instead of the other is scheduled. 
In the example, the job is scheduled due to minor run time ($f_5$) and task replicates ($f_6$) of its smallest ready node.

\begin{table}[t]
\centering
\small
\caption{Features of the paired jobs. The stared job is scheduled.}
\setlength{\tabcolsep}{0.3mm}{
\begin{tabular}{lcccccccr}
\hline
$Job_1$ (*) & 112.44 & 614 & 115.88 & 616 & 4.62 & 2 & False \\
$Job_2$ & 1385.17 & 1080 & 1600.87 & 1186 & 215.7 & 580 & False \\ \hline
\end{tabular}}
\label{tab:example_job_pair}
\end{table}

\begin{figure}
    \centering
    \includegraphics[width=0.75\linewidth]{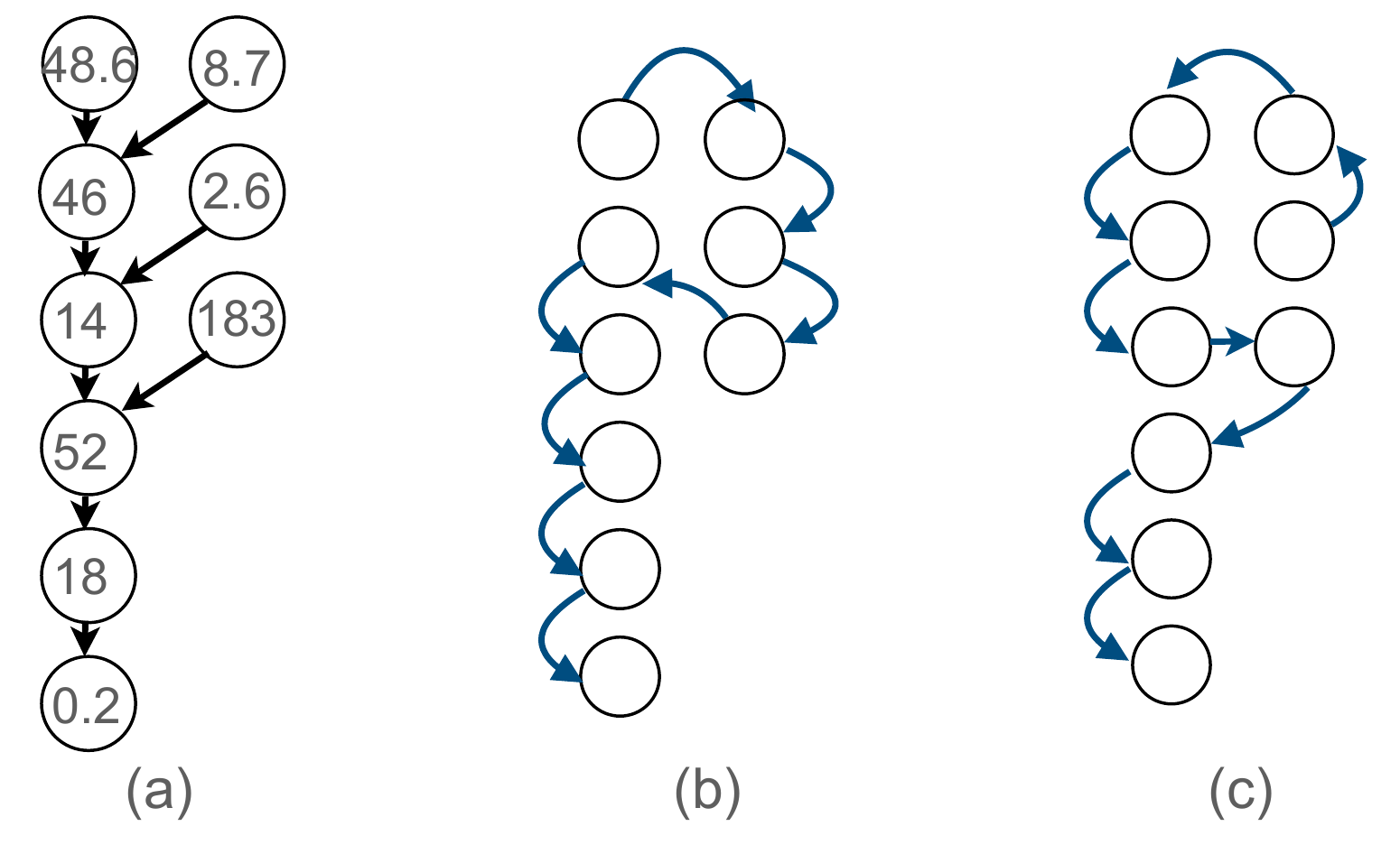}
    \caption{(a) The job detail, e.g. run time of and dependencies between task nodes, and the task schedule based on  (b) FCFS and (c) SNF rules. Arrows in (a) show dependencies, while in (b) and (c) show the execution order. The dependencies in (b) and (c) are the same with (a), and are omitted for clearance.}
    \label{fig:task_level_explanation}
\end{figure}

\paragraph{Task level explanation} The $g_n$ provides an understanding of how the task is scheduled within a job by aligning the DRL-based policy with prevalent rules. As shown in Figure~\ref{fig:explanation-decima}, the policy diverges on different branches of the tree, aligning to different rules. For the job selected by locality (on the left branch), the FCFS rule provides the highest approximation accuracy of 0.71. While for decision explained with random forest (on the right branch), the SNF rule achieves the highest approximation accuracy of 0.83. We explain with an example how the task is scheduled differently when the rules apply. As shown in Figure \ref{fig:task_level_explanation}, the job contains 9 task nodes, of which the estimated run time and dependency are shown on the left. The task-level schedule when it is local to the executor (middle) and when it is not (right) provides clear logic and an understandable pattern for the administrator.

\paragraph{Overall understanding} We obtain the following understanding from the above explanations for Decima's policy. In most cases (89\%), the job local to the free executor is scheduled. The tasks within the job are scheduled largely (71\%) based on the FCFS rule. In other cases when no job is local to the free executor, the logic behind the policy can be revealed through a comparison between the scheduled job and those not scheduled with high faith (0.88). Within the job, tasks are largely (83\%) scheduled based on the SNF rule. 

\begin{figure}[t]
    \centering
    \begin{minipage}[t]{0.46\linewidth}
    \centering
    \includegraphics[width=0.95\linewidth]{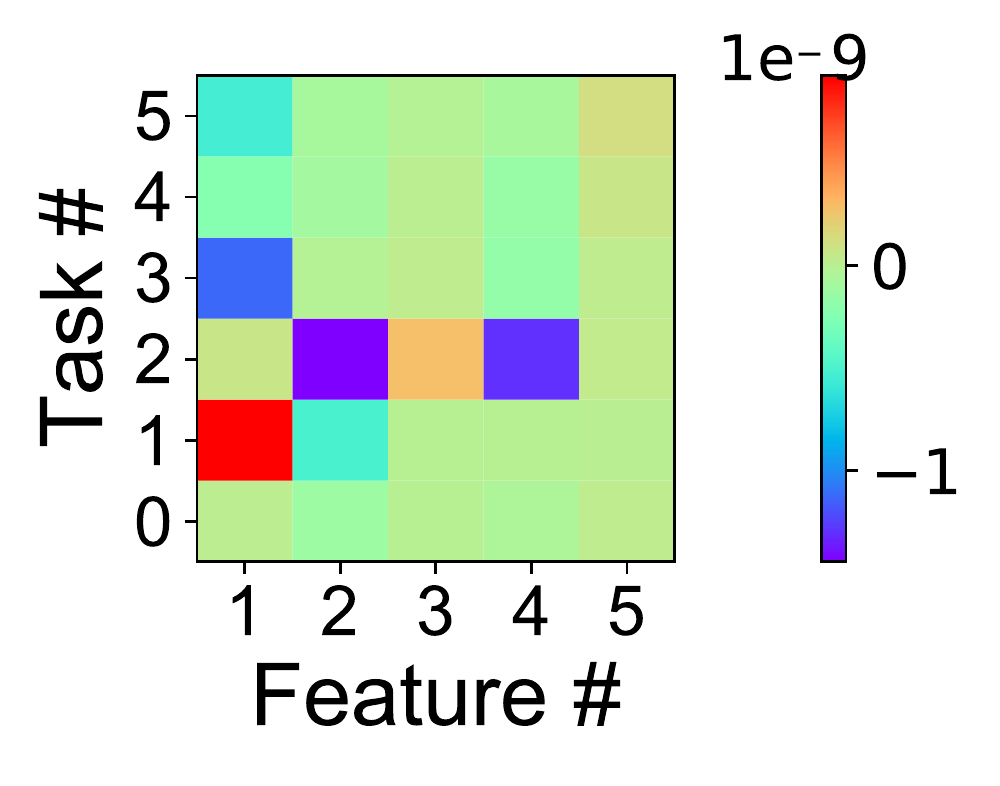}\\
    \centering{\small{\text{(a)}}}
    \end{minipage}%
    \hspace{1.2mm}
    \begin{minipage}[t]{0.46\linewidth}
    \centering
    \includegraphics[width=0.95\linewidth]{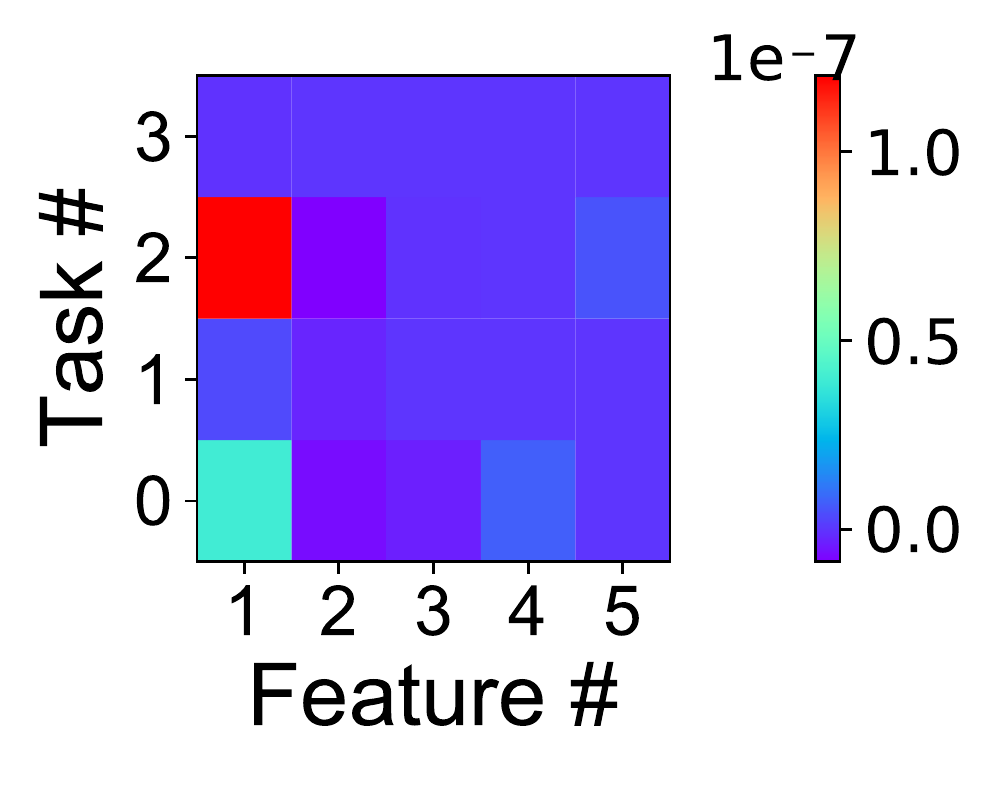}\\
    \small{\text{(b)}}
    \end{minipage}%
    \caption{The salience map of (a) the positive job and (b) the negative job. The positive job is scheduled at the current time stage. }
    \label{fig_baseline}
\end{figure}

\paragraph{Explanation compared with baseline.} We show how the same job decision is explained with the saliency-based approach. The salience map is formed over the original feature space of all 30 jobs and cropped based on two aforementioned jobs paired and explained with the RF model in Figure \ref{fig:rf_explanation_examples}. As shown in Figure \ref{fig_baseline}, the salience values are mostly close to 0, meaning they are not prominent in the DRL-based scheduling decision. However, it still figures out the relative importance of task features through comparison. For example, the $Feature~ 1$, namely, the executors owned by $Task~1$ of the positive job and $Task~2$ of the negative job, has a greater impact. However, the explanation is too elementary and less intuitive than what is shown by the multi-level explanations.      

\subsection{Implication}
Understanding the DRL-based schedule behavior is helpful to analyze its robustness issue. Decima shows the following behavior pattern that favors smaller tasks, according to the interpretable feature 5 and SNF rule. This pattern may be made use of by a specific job owner. We show that this behavior has some robustness issues and a user can get benefits (e.g. faster completion) if he knows the pattern. 

\begin{figure}
    \centering
    \begin{minipage}[t]{0.5\linewidth}
    \centering
    \includegraphics[width=0.95\linewidth]{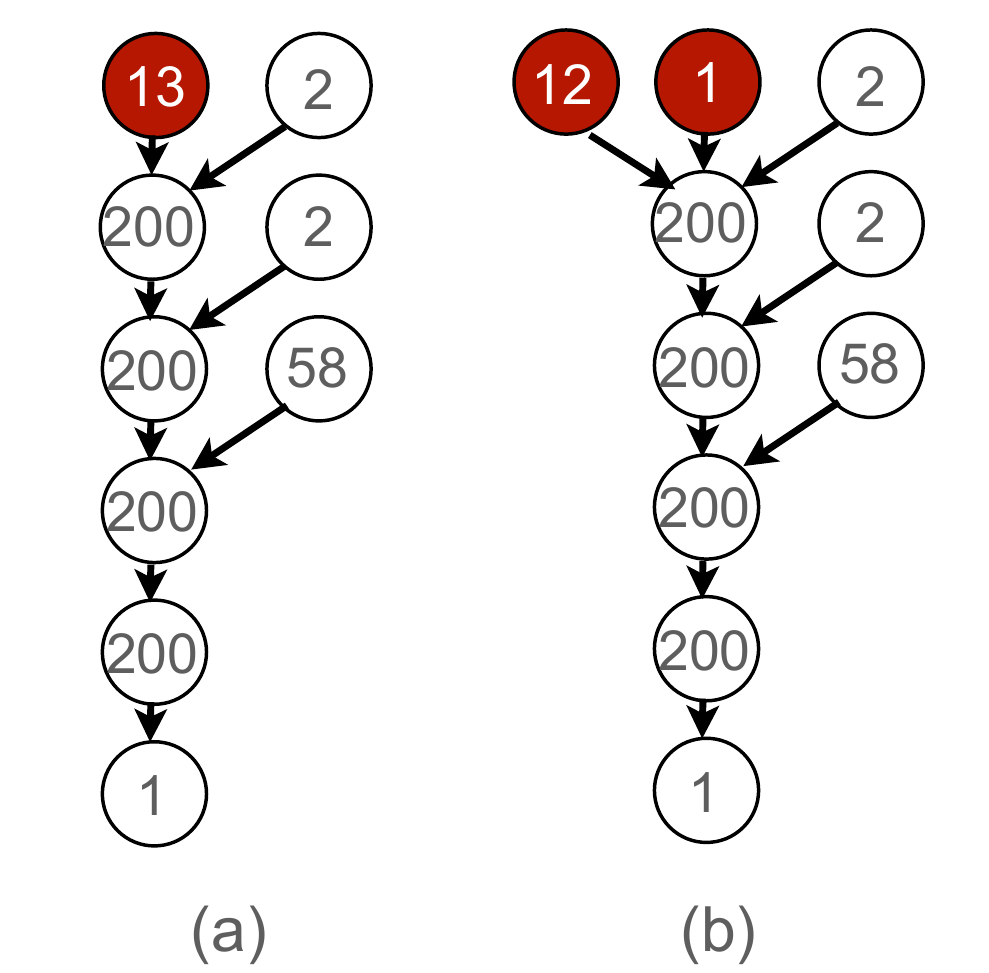}
    \end{minipage} %
    \begin{minipage}[t]{0.47\linewidth}
    \centering
    \includegraphics[width=0.85\linewidth]{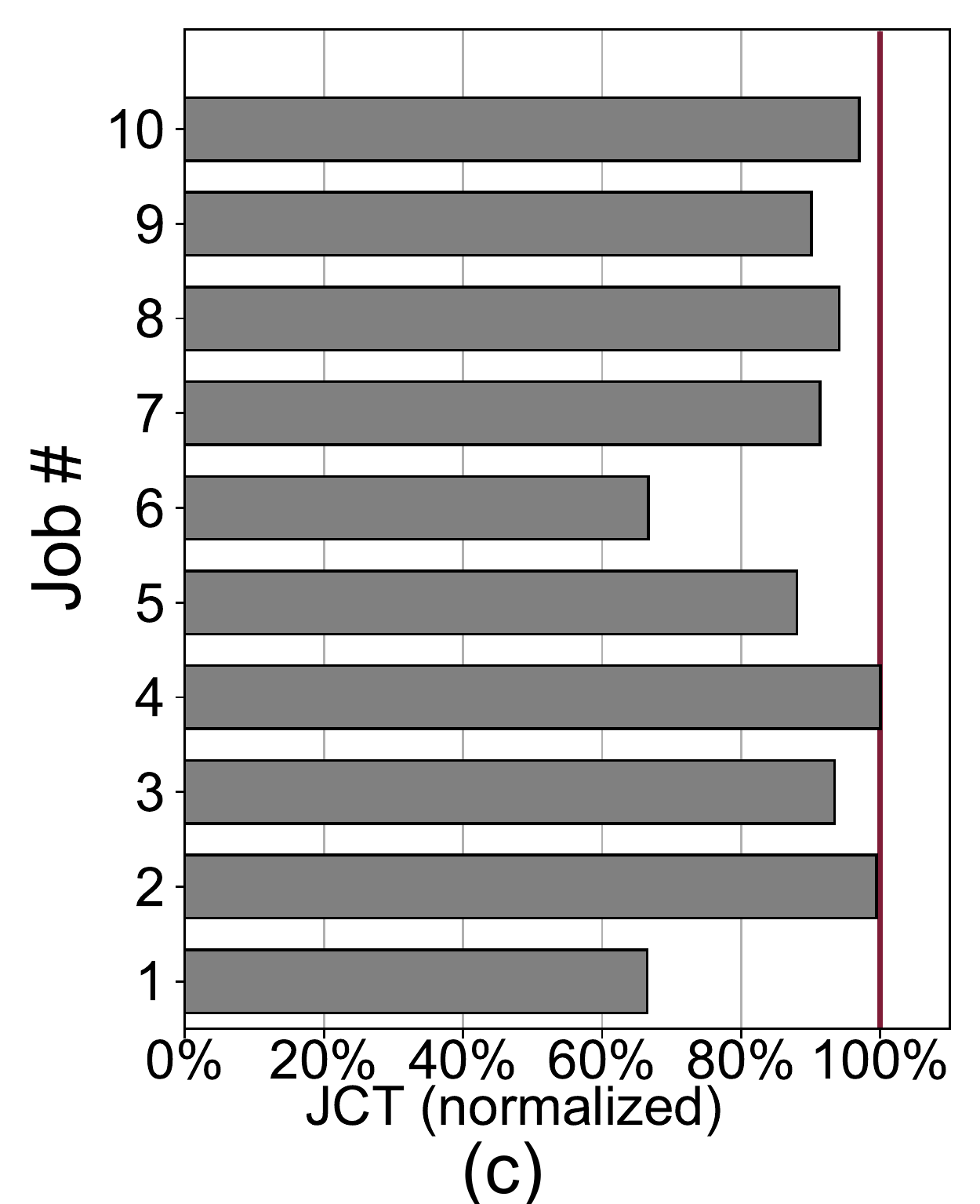}
    \end{minipage}
    \caption{The example of \emph{node-split} perturbation on a TPC-H job DAG: (a) the original job and (b) perturbed job. Red color shows the task node for split. (c) The JCT of perturbed job normalized to the original job. Less than 100\% means faster completion of perturbed jobs. All jobs except $Job~4$ are benefited.}
    \label{fig:exemplary_job_dag_perturbation}
\end{figure}

We devise the \emph{node-split} perturbation to adjust the job. Since the task node contains replicates of the same codes running on different data in parallel, it is feasible to split the data and generate two smaller nodes instead of the original one. Take Spark as an example, splits on the RDD inputs of a computational stage change the node to two new nodes, each running on part of its original data. An example of a TPC-H job and its perturbation are shown in Figure~\ref{fig:exemplary_job_dag_perturbation}. The number of task replicates is included in each node. Originally, the shaded task node contains 13 replicates as shown in Figure~\ref{fig:exemplary_job_dag_perturbation}(a). It is split into two new nodes containing 1 and 12 tasks respectively, as shown in Figure~\ref{fig:exemplary_job_dag_perturbation}(b). The two jobs are executed with the same collection of background jobs independently. All jobs are scheduled by Decima. Results show that the perturbed job is completed 8.7\% faster.

We conduct experiments on more jobs and find it is not a single case. Basically, 10 jobs are crafted, each running in two versions either (1) without perturbation or (2) with perturbation. They are executed with the same collection of background jobs and the job completion time (JCT) is recorded. Then, the JCT of the perturbed version is normalized to that of the original job. As shown in Figure~\ref{fig:exemplary_job_dag_perturbation}(c), 9 jobs are completed faster (with normalized JCT $<$ 100\%), and the average benefit on JCT is 11.3\%. The reason behind this is that Decima opts to schedule tasks with less run time. After the node split, the smaller task has a chance to be scheduled earlier. Due to the job's locality to the executor, the remaining tasks in this job will have a greater chance to reuse this executor and introduce a shorter overall completion time. However, it is not fair if the executor is occupied by this job only. It may induce an unexpected slowdown in the background jobs. For a system administrator, understanding the behavior pattern is important to make operational plans to manage such potential problems. 

\section{Conclusion}
Deep reinforcement learning-based schedulers show remarkable performance improvement in dependency-aware job scheduling despite its hard-to-interpret decision-making process. In this paper, we demonstrated a multi-level explanation framework from the perspective of system administrators to achieve simplicity in system management. It could close the gap between the theory and application. We evaluated the performance of our framework on the TPC-H benchmark to explain a complex DRL-based scheduler. Our method provided both accurate approximation and meaningful explanation. The obtained insights helped to identify the vulnerability of the DRL-based scheduler. Overall, our work established a pathway for the complex DRL model to gain the trust of the system administrator.

\end{document}